\documentclass[12pt]{article}
\usepackage{times}
\usepackage{geometry}
\usepackage[utf8]{inputenc}
\usepackage{enumitem,amssymb,graphicx,hyperref}
\usepackage{natbib}

\usepackage{aas_macros}

\usepackage{ragged2e}
\newlist{thematic}{itemize}{8}
\setlist[thematic]{label=$\square$}
\usepackage{pifont}
\newcommand{\cmark}{\ding{51}}%
\newcommand{\done}{\rlap{$\square$}{\raisebox{2pt}{\large\hspace{1pt}\cmark}}%
\hspace{-2.5pt}}

\bibpunct{(}{)}{;}{a}{}{,}
\begin{document}
\RaggedRight
\huge
Astro2020 APC White Paper 

\vskip 8pt 
Next Generation LSST Science  \linebreak
\normalsize

\noindent \textbf{Type of Activity:} \hspace{0.5cm}
$\done$ Ground Based Project \hspace{0.5cm} 
$\square$ Space Based Project \linebreak 
$\square$ Infrastructure Activity \hspace{0.5cm}
$\square$ Technological Development Activity \linebreak
$\square$ State of the Profession Consideration \hspace{0.5cm}
$\square$ Other \linebreak

\noindent \textbf{Thematic Areas:} \hspace*{60pt} $\done$ Planetary Systems \hspace*{10pt} $\done$ Star and Planet Formation \hspace*{20pt}\linebreak
$\done$ Formation and Evolution of Compact Objects \hspace*{31pt} $\done$ Cosmology and Fundamental Physics \linebreak
  $\done$  Stars and Stellar Evolution \hspace*{1pt} $\done$ Resolved Stellar Populations and their Environments \hspace*{40pt} \linebreak
  $\done$    Galaxy Evolution   \hspace*{45pt} $\done$             Multi-Messenger Astronomy and Astrophysics \hspace*{65pt} \linebreak
  
\textbf{Principal Author:}

Name:	Saurabh~W.~Jha
 \linebreak						
Institution:  Rutgers, the State University of New Jersey
 \linebreak
Email: saurabh@physics.rutgers.edu
 \linebreak
Phone:  +1 848-445-8962
 \linebreak
 
\textbf{Co-authors:} 
  \linebreak
Federica Bianco, University of Delaware \\
W. Niel Brandt, Pennsylvania State University \\ 
Gaspar Galaz, Pontificia Universidad Cat\'{o}lica de Chile  \\ 
Eric Gawiser, Rutgers University \\
John Gizis, University of Delaware \\ 
Ren\'{e}e Hlo\v{z}ek, University of Toronto \\
Sugata Kaviraj, University of Hertfordshire \\ 
Jeffrey A. Newman, University of Pittsburgh  \\ 
Aprajita Verma, University of Oxford \\ 
W. Michael Wood-Vasey, University of Pittsburgh \\

\vskip 0.5in
\textbf{Abstract:}
The Large Synoptic Survey Telescope (LSST) can advance scientific frontiers beyond its groundbreaking 10-year survey. Here we explore opportunities for extended operations with proposal-based observing strategies, new filters, or transformed instrumentation. We recommend the development of a mid-decade community- and science-driven process to define next-generation LSST capabilities.

\pagebreak

\justifying
\parindent=0pt
\parskip=10pt

\section{Key Science Goals \& Objectives}

The highest priority large-scale ground-based project from the Astro2010
decadal survey, the Large Synoptic Survey Telescope (LSST), is planned to
undertake a 10-year survey beginning in 2022 \citep{LSST:2009,Ivezic:2019}.
LSST and its 3.2 gigapixel camera will image the southern sky in six optical
filters with an unprecedented combination of area, depth, and cadence,
enabling a huge range of science. While the details of the observing strategy
for the 10-year survey and its components (including the wide-fast-deep main
survey, deep-drilling fields, and other mini-surveys) are still being refined,  
now is the time to start planning what comes after the 10-year
survey and considering the next generation of LSST science. This connects to
\emph{all} of the science thematic areas and numerous science-focused White
Papers.

Here we consider a selection of science cases that would motivate three 
scenarios for next generation LSST operations: 
\begin{enumerate}
    \item No modifications to the telescope or camera, but complete flexibility to undertake new observing strategies, cadence, sky coverage, exposure time, etc., building on the discoveries and scientific progress made during LSST operations and driven by the global scientific landscape at the completion of the 10-year survey.
    \item 
    Modest modifications to the camera, e.g., new filters.
    \item Replacement of the camera with another instrument (e.g., a wide-field multiplexed spectrograph or near-infrared camera) or other major modifications.
\end{enumerate}
We describe these possibilities in more detail in the following sections.

\subsection{LSST + LSSTCam: Continuing Operation of a Flagship Facility}
\label{sec:carryon}

LSST emerged as the highest-ranked large-scale ground-based project in the
Astro2010 Decadal report \citep{NAP12951}, as was reiterated in the Midterm
Assessment \citep{NAP23560}, because of its ``transformational'' scientific
impact. As such LSST will be a flagship ground-based observatory and there is
every reason to expect the telescope and camera will continue to be
scientifically valuable ``as is'' well beyond the envisioned 10-year
survey. 

Even with the same instrumentation, extended operation of LSST will open up a
wide range of observing strategies with more flexibility than is planned for
the 10-year survey. Even working within the 10-year survey constraints, the
2018 LSST call for observing strategy white papers
(\url{https://www.lsst.org/call-whitepaper-2018}) yielded 46 submissions
(\url{https://www.lsst.org/submitted-whitepaper-2018}).  The papers included
varying choices for cadence in the main 10-year ``Wide-Fast-Deep'' (WFD)
survey, proposed deep-drilling fields (DDFs), and target-of-opportunity (TOO)
observations. More than half of the submitted papers proposed ``mini-surveys''
with a wide variety of scientific aims and scope of observational resources
required. The proposals easily oversubscribed the 10-year limit, by about a
factor of two in total \citep{Jones:2019}, and the oversubscription of the
non-WFD observing time is much higher.

The LSST Project is undertaking a process to optimize and iterate the
observing strategy to respond to these requests \citep{LSST-SAC}, but it is
clear that not all of these ideas can be accommodated in the 10-year survey.
Moreover, the white papers were submitted \emph{assuming 10-year survey
constraints}; the flexibility afforded by extended LSST observations will
surely allow for more novel ideas varying depth (exposure time), sky area,
filters, and cadence.

Opening extended LSST operations to a proposal-based observing strategy has
several advantages, the foremost of which is that it allows the most
compelling science to drive the observations, without needing to know years in
advance what that highest-priority science will be. A proposal-based process
can also accommodate concurrent scheduling of surveys with different
timescales, ranging from a fraction of a night to multiple years of data.

``Static-sky'' science that makes use only of the 10-year co-added LSST images
over a large fraction of the sky (i.e., with metrics that scale as $\sqrt{t}$)
may not benefit greatly from extended operations without a long time baseline.
However, ``dynamic-sky'' science, ranging from the Solar System to cosmology,
would continue to provide opportunities which cannot be matched at other
facilities.

\begin{figure}
\begin{center}
\includegraphics[width=\textwidth]{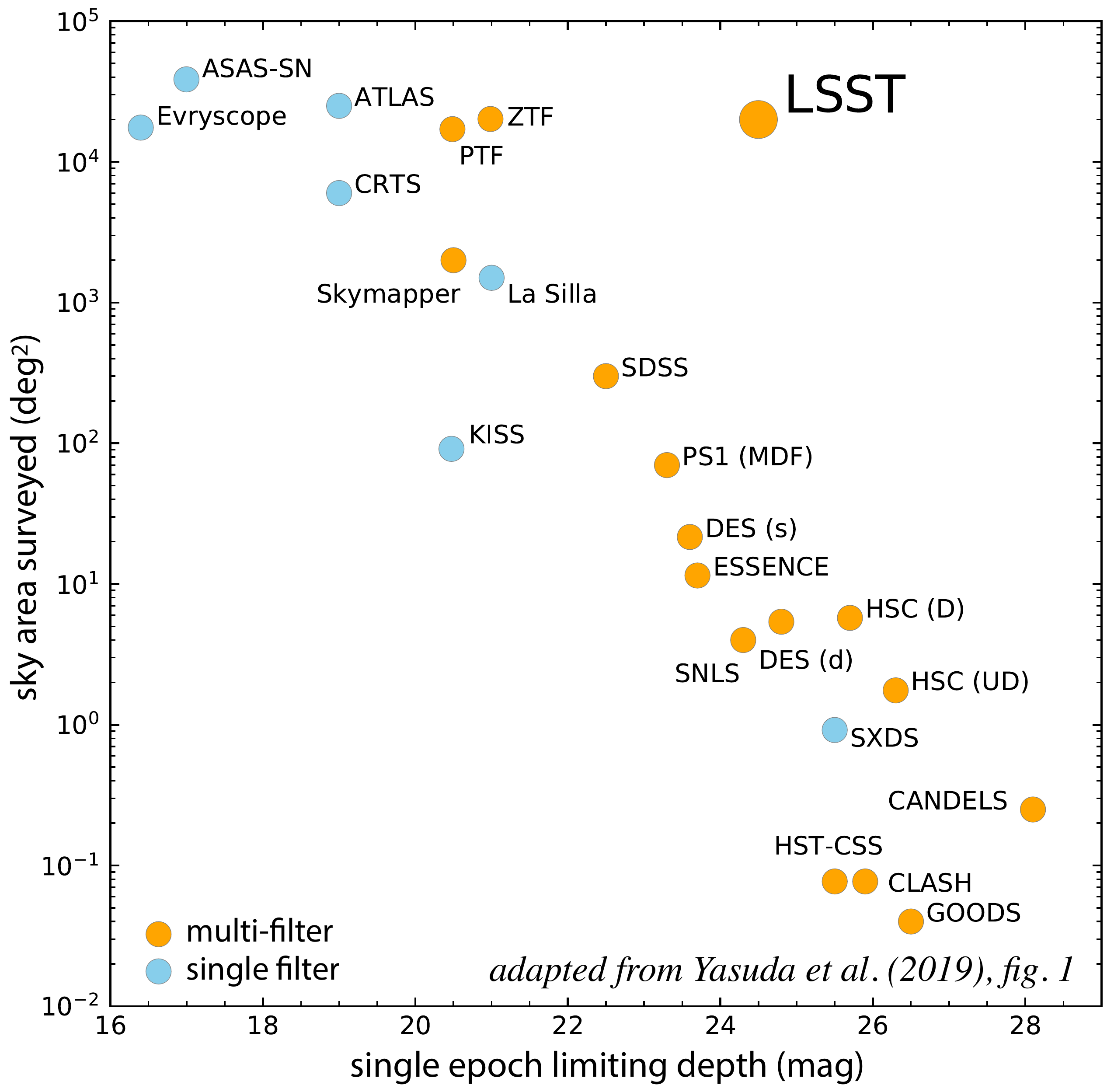}
\vskip -0.1in \caption{Sky area versus single-epoch depth for time-domain
optical surveys. LSST is unique in its ability to characterize the changing
sky, and it should remain as the premier facility in this application even
after its 10-year survey. \emph{This figure is adapted from
\citet{2019PASJ..tmp...74Y}.} \
 \label{fig:surveys}}
\end{center}
\end{figure}

Figure~\ref{fig:surveys} shows the unique position LSST will occupy in
exploring the optical time-domain \citep{2019PASJ..tmp...74Y}. LSST will play
the starring role in the era of celestial cinematography, but the movie of the
Universe does not end in 2032. We can be confident that time-domain optical
astronomy will remain a compelling field through the 2030s and beyond; if LSST
ceases to be the chronicler of the changing sky, another facility would need
to take its place. Furthermore, new large multiwavelength/multimessenger
facilities beginning operations in the 2030s require simultaneous or
near-simultaneous LSST observations. For example, studies of fast
extragalactic X-ray transients discovered by the large-grasp Athena X-ray
Observatory \citep[e.g.,][]{Yang:2019} will benefit greatly from supporting
LSST observations, as would studies of fast radio transients from the SKA, and
gravitational wave events from LIGO Voyager, Cosmic Explorer, and LISA.

\subsection{Upgrading LSSTCam: Building on an LSST Legacy}
\label{sec:minor}

Alternatively, while keeping LSST focused on wide-field optical imaging, minor
modifications to the camera could have a large scientific impact at modest
cost. One of the simplest to consider is expanding the LSST filter set beyond
the currently planned six filters (\textit{ugrizy}). Additional filters would
not preclude continued use of the original filters (though having only five
slots in the filter exchanger may become more of an issue), and as such would
complement the increased flexibility in observing strategy described above.

\citet{2019BAAS...51c.303Y} describe a science case for 15-20 nm FWHM
narrow-band filters for LSST, with applications to deriving physical stellar
parameters, mapping emission-line nebulae, improving galaxy photometric
redshifts, and creating cosmologically-useful samples of extragalactic objects
in narrow redshift shells. Surveys with narrow-band filters are not well
suited for the majority of transients and variables, so extensive time-domain
coverage or cadenced observations are not essential, and thus large sky area
narrow-band images (to scientifically interesting depths) can be obtained
economically, in just nights to weeks of survey time.

A set of narrow-band filters with widths of 15--20 nm (the smallest allowed by
the \textit{f}/1.2 beam), would build on the results of a variety of past
projects which used $\sim$30 nm FWHM medium-band filters with good results
\citep[e.g.,][]{Wolf:2004,Ilbert:2009,Cardamone:2010}. Implementing a
carefully-designed narrow-to-medium-band filter set for LSST could enable
greatly improved photometric redshifts over limited sky areas (in contrast
photo-\textit{z} uncertainties would only be modestly improved if the
observations cover the full LSST sky footprint in the same survey time).  A
particular opportunity would be presented by surveys covering regions with
grism observations from WFIRST, in which case effective low-resolution
spectroscopy of hundreds of millions of objects covering a broad wavelength
range in the optical and near-infrared would be available. Even considering
LSST photometry alone, a 10-year survey over 2000 square degrees (instead of
20,000) with 30 optical bands would yield photo-\textit{z} uncertainties at
least a factor of 7 smaller than the original LSST survey\footnote{In the
limit of constant number of effective spectral features, equal filter
coverage, and fixed sky area, photometric redshift uncertainties should scale
as $N_{\rm bands}^{-1/2} \; t_{\rm total}^{-1/2}$, where $N_{\rm bands}$ is
the number of photometric passbands and $t_{\rm total}$ is the total observing
time.  If effective spectroscopic resolution increases significantly, new
spectral features start to be accessible, making this a worst-case
scaling.}(i.e., redshift uncertainties for \textit{i} $\sim$ 25 galaxies of
$\sim$0.005(1 + \textit{z}), with smaller uncertainties for brighter objects),
which would have high impact across many science areas (particularly studies
of galaxy evolution or WFIRST cosmological measurements).

A ``shifted'' broad-band filter set (i.e., with the same filter widths, but
shifted central wavelengths, for example halfway between the current filter
centers) would provide additional opportunities. Such filters would still have
high throughput for single-visit discovery (transients, variables, moving
objects, etc.) and could be optimized to improve estimates of galaxy
properties including photometric redshifts and/or measurements of stellar
parameters. A 10-year survey with shifted filters would reduce photometric
redshift uncertainties across the full sky by $\sim\sqrt{2}$ and provide
greater effective spectral resolution on all objects for astrophysical
studies, while continuing to provide discoveries and light curves for
transients at a comparable rate to the original LSST survey.

New filters could be developed where the bandpass varies over the focal plane.
The variation might be continuous as a function of position on the detector,
or else define discrete regions \citep[e.g., like the SDSS imaging
camera;][]{Gunn:1998}. This could enable rapid tiling of the sky in multiple
wavelengths, leveraging the fast slew time of LSST (seconds) compared to the
slower filter change time (minutes). Cleverly patterned bandpasses (e.g., a
checkerboard of square tiles) combined with an appropriate dither pattern
could allow contiguous sky coverage in all of the bandpasses without any
filter changes. Such an approach could also allow a larger complement of
passbands to be available to LSST at all times, at the expense of the field of
view for any given passband. Photometric calibration of these variable filters
could be challenging, but variation of the response across the focal plane
must already be calibrated for the ``uniform'' filters \citep{Ivezic:2019}.

Polarizing filters for LSST should also be investigated. These could allow
studies of interstellar magnetic field structure, asymmetries in explosive
transients, jets of radio-loud active galaxies, and obscuration geometry in
active galaxies.

\subsection{New Instrumentation for a Next Generation Observatory}
\label{sec:major}

Major upgrades to the telescope and its instrumentation could transform LSST
into a truly ``next generation'' observatory. Approaches along these lines
need to consider both what would complement the LSST data set after the
10-year survey as well as the landscape of other facilities in the 2030s and
beyond.

Some types of advanced instrumentation would keep LSST capable of wide-field
optical imaging, but with enhanced capabilities. This might minimally include
maintenance and repair of LSSTCam or, more substantially, replacement with
another optical camera. Beyond this, upgrades might improve the image quality
through active optics and/or shifting charge on the detector. Similarly,
utilizing energy-sensitive detectors like MKIDs \citep{2018SPIE10702E..0HM}
could enable both higher overall throughput and observing efficiency
\emph{and} higher spectral resolution than broad-band filters. All of these
more speculative ideas would require significant technological development and
should be explored.

Alternately, LSST could radically shift from wide-field optical imaging to
other modes, most notably wide-field near-infrared imaging or wide-field
multiplexed spectroscopy in the optical and/or infrared. The science cases for
large surveys in both of these modes are well developed \citep[e.g.,]
[]{2019BAAS...51c.296K,2019BAAS...51c.360C,2019BAAS...51c.418E,2019BAAS...51c..30R,2014SPIE.9147E..0MD,2014SPIE.9147E..0SF,2018arXiv181008695H,2019BAAS...51c.363M}.
The main question would be to identify an important niche that could best be
filled by a transformed LSST. Competition for wide-field near-infrared imaging
may come from space-based facilities with better spatial resolution ( e.g.,
Euclid and WFIRST). Spectroscopic surveys that stare at limited fields with
long exposure times would not take advantage of the fast-slewing capability of
LSST.

The need for very wide-field spectroscopy with LSST-like telescope aperture to
exploit the rich LSST imaging dataset has been repeatedly identified
\citep[e.g., in][]{2016arXiv161001661N}; if no route to such spectroscopy is
secured by the end of the 2020s, it may be necessary to explore how LSST can
be used in this role after its 10-year survey is completed. As mentioned
above, the opportunity cost of eliminating a leading wide-field optical imager
for time-domain science should also be considered, and some science cases
(particularly in the time domain) require spectroscopic observations
contemporaneous with LSST optical imaging \citep[see also,
e.g.,][]{2019BAAS...51c.369H}.

\section{Technical Overview}

Technical considerations associated with extended operation of LSST in these
three scenarios will be presented in a separate white paper by the LSST
Project. \citet{2019arXiv190504669S} report on the LSST Next Generation
Instrumentation workshop held in April 2019, and note the scientific benefits
and technical challenges associated with new LSST filters, new detector
technology, wide-field multiplexed spectroscopy, and wide-field near-infrared
imaging.

\section{Technology Drivers}

The more radical changes to LSST instrumentation discussed here would require
significant technology development. Improvements to the image quality in the
optics or detector, energy-sensitive detectors, and concepts for wide-field
multiplexed spectroscopy or near-infrared imaging all need exploration.

\section{Organization, Partnerships, \& Current Status}

The LSST 10-year survey is jointly supported by the US NSF and DOE, and the
science community comprises the United States, Chile, and other select foreign
partners. A funding model for extended operations would need to be developed.

\section{Schedule}

Decisions about the future of LSST after the 10-year survey should be
community and science driven. We recommend that the Astro2020 Decadal Survey
help develop a timeline for this process, taking into account the context of
other facilities that should come online in the 2020s (ELTs, JWST, wide-field
ground- and space-based imaging and spectroscopic capabilities, etc.) as well
as likely future scientific priorities. The 2025 Midterm Assessment may
provide an opportune time to make concrete recommendations for the future of
LSST. We advocate a review after the first LSST full-year data release
(c.~2024) to evaluate what the most productive uses for LSST may be in the
2030s, taking into consideration the real-world performance of LSST, the
optical/infrared astronomical landscape, and the evolving landscape of the
most compelling scientific directions.

\section{Cost Estimates}



A major upgrade to LSST instrumentation (\S~\ref{sec:major}) would almost
certainly be in the Large ground-based cost category ($>$\$70M). Even minor
(\S~\ref{sec:minor}) or no (\S~\ref{sec:carryon}) instrumentation changes
would still require significant annual operations costs for the observatory,
but would provide excellent added value to the investment in telescope and
camera construction and the first ten years of operations, building on the
developed hardware, computing, and science infrastructure. Additional costs
would be associated with a proposal process for observing strategies.
Specialized data reduction or analysis pipelines may be needed for some
proposed observations and would need to be supported. Depending on the
timescale for extended operations (e.g., several years), this again could
reach the Large ground-based cost category. More details regarding the cost of
LSST extended operations are provided by the LSST Project Team white
paper.

\section{References}

\bibliography{lsstng}

\end{document}